\documentclass[10pt, twocolumn]{article}
\usepackage[utf8]{inputenc}
\usepackage[english]{babel}
\usepackage{amssymb}
\usepackage{amsmath}

\usepackage{lineno}
\usepackage[margin=0.75in, top=0.75in]{geometry}

\usepackage{algorithm}
\usepackage[noend]{algpseudocode}

\usepackage{float}

\floatstyle{boxed}
\newfloat{deffloat}{h}{deffloats}
\floatname{deffloat}{Definitions}

\usepackage{empheq}

\usepackage{framed} % or, "mdframed"
\usepackage[framed]{ntheorem}
\theoremstyle{nonumberplain}
\newframedtheorem{frm-thm}{}

\usepackage{multicol}
\usepackage{sectsty}
\sectionfont{\bfseries\Large\raggedright}
\sectionfont{\bfseries\large\raggedright}

\usepackage{enumitem}
\newlist{todolist}{itemize}{2}
\setlist[todolist]{label=$\square$}

\usepackage{xifthen}
\newcommand{\E}[2][]{%
  \ifthenelse{\isempty{#1}}{%
    \mathrm{E}\left[#2\right]
  }{%
    \mathrm{E}\left[#2 \middle| #1\right]
  }%
}

\newcommand{\Cov}[2][]{%
  \ifthenelse{\isempty{#1}}{%
    \mathrm{Cov}\left(#2\right)
  }{%
    \mathrm{Cov}\left(#2, #1\right)
  }%
}

\usepackage{graphicx}
\graphicspath{ {./images/} }
\usepackage[export]{adjustbox}
\usepackage{subcaption}

\usepackage[
backend=biber,
style=alphabetic,
citestyle=authoryear
]{biblatex}
\usepackage[colorlinks,citecolor=blue,urlcolor=blue,bookmarks=false,hypertexnames=true]{hyperref}

\author{
  Jeffrey Wong\\
  Apple
  \and
  Jasmine Nettiksimmons\\
  Apple
  \and
  Jiannan Lu\\
  Apple
  \and
  Katherine Livins\\
  Apple
}

\title{Addressing Hidden Imperfections in Online Experimentation}
\date{}

\begin{document}
\maketitle

\section{Introduction}

Randomized controlled trials (RCTs) have long been the gold standard for causal inference in scientific fields, particularly in biomedicine (\cite{rubin2008objective, imbens2015causal}). However, it is well known that in many cases the ideal RCT cannot be conducted. For example, some patients in medical trials pass away or become otherwise untrackable (\cite{rubin2006causal}), and others may not fully comply with their assigned treatments (\cite{efron1991compliance, frangakis2002principal, jin2008principal}).

Technology companies are increasingly using RCTs as part of their development process. Despite having fine control over engineering systems and data instrumentation, these RCTs can still be imperfectly executed. In fact, online experimentation suffers from many of the same biases seen in biomedical RCTs including opt-in and user activity bias (\cite{wang2019heavy}), selection bias (\cite{dasgupta2012overcoming, dmitriev2016pitfalls, xie2021measure}), non-compliance with the treatment (\cite{deng2015diluted}), and more generally, challenges in the ability to test the question of interest (\cite{ gupchup2018trustworthy}). The result of these imperfections can lead to a bias in the estimated causal effect, a loss in statistical power, an attenuation of the effect, or even a need to reframe the question that can be answered.

This paper aims to make practitioners of experimentation more aware of imperfections in technology-industry RCTs, which can be hidden throughout the engineering stack or in the design process. We recommend designers of experiments to be vigilant and iterate together with product and user experience designers to reconcile learning goals with minimization of burden on end consumers.  We will demonstrate the need for mature thought with a recurring example. As we describe the challenges, we will discuss what is possible to infer from a RCT using a difference of means, and whether the original business question can be answered. Later, we will show an analytics strategy to minimize any gaps between the two. We offer practical guidance on how to 

\begin{enumerate}
\item Design and scope the experiment.
\item Instrument the experimentation funnel.
\item Proactively monitor imperfections in measurement.
\item Adjust statistical analysis of an experiment to mitigate imperfections.
\end{enumerate}

These concepts will be illustrated with a running example that assumes on-device treatment assignment. This scenario has broad challenges that many practitioners may be unfamiliar with. Server-side experiments suffer from some of the same imperfections and can also benefit from these guidelines.

\section{Designing and Scoping the Experiment}

%Despite the ease that experimentation platforms provide in creating, launching, managing and analyzing experiments,
Experimenters need to be vigilant and thoughtful in experiment design. Conclusions from experiments can suffer from opt-in bias, selection bias, non-compliance with the treatment, or effect attenuation. To start a rigorous experiment, we must first frame a business question with an appropriate target. Afterwards, the designer of the experiment must think carefully about

\begin{enumerate}
    \item How will the users’ experience be randomized?
    \item How will a user trigger that randomized experience?
    \item What is the broader target population, and how will a subset of users enter the experiment?
    \item Are there any mechanisms that create an unequal randomization in the treatment and control experiences?
\end{enumerate}

Multiple iterations across teams may be required on these questions. Experiment design, user experience design, and business goals can conflict with each other, and the need to balance them may lead to experiments that are difficult to conclude. Natural limitations in engineering and design can create an imperfect experiment that affects the scope of the conclusion, and the generalizability of the results.

First, we must start with a question. Throughout this paper we will rely on a recurring example about an online store that sells products in a mobile app. The product pages have images, and the store has released an update to their app that enables the product pages to include video. The question is simply “How does the new video for product X affect sales?” 

%The implementation for the videos requires the user to be on the latest version of the app (\cite{xu2016evaluating}). If this feature is enabled, the app will pull a configuration that arranges the product page, but this configuration may fail to download. When a user lands on a product page, the user will see a video and have the opportunity to play it. %Below we will show how limitations in randomizing, triggering, and entering the experiment affect the scope of the conclusion.

Given this background, the designer of the experiment must design a randomized controlled trial to understand the causal effect of the video on sales. The naive design is to randomly show the video to a fraction of the online traffic, then compute sales metrics for users that watched the video and those who did not watch the video. The analysis would take the difference in the means of sales and report that as the causal effect of the video. While this seems like a reasonable experiment, \textbf{there are many hidden imperfections that affect the scope and validity of the results}. It is the designer’s duty to state these imperfections, and to scope the results of the experiment clearly.

Imperfections arise when studying how a user gets to see the video. It is discovered that a series of events must happen:

\begin{enumerate}
    \item The user must be on the latest version of the app.
    \item If the video feature is enabled, the app must download a configuration onto the device, allowing it to arrange the product pages. It is possible that this download fails.
    \item The user must land on the product page for X.
    \item On the product page, the user will see a video player. They must click on it in order to play it.
\end{enumerate}

Among the imperfections are (1) the possibility that users do not play the video, (2) selection bias due to the treatment configuration failing to download, while the control is configured without failure, and (3) changes in the types of users that land on the product page. Simply computing a difference in means for everyone that was in the experiment, regardless of playing the video, will decrease statistical power, and simply removing users assigned to the treatment that did not watch the video, or could not download the treatment configuration successfully, will result in bias.
In the next sections, we will show how randomizing, triggering, and entering the experiment affect the scope of the conclusion.
Planning and analytical support are needed to address the imperfections and have a robust conclusion. 

\subsection{Randomizing the Experiment}

%We may want to define the treatment group as users who have played the video, and the control group as users who did not not play the video.
We are tempted to report the effect of the video by taking the difference between users who have played the video, and users who did not.
However, this is not a randomized partition of users that the experiment designer can create apriori. Instead, the app can only use randomization to control the availability of the video. We define this as the intent to treat (ITT), which can be reported using a difference in means without bias as long as the randomized assignment was logged independently of whether the user successfully downloaded the video configuration. When analyzing the ITT effect, the scope of the conclusion should be qualified to “the effect of making the video available is ...”, instead of “the effect of watching the video is ...”. The designer should clearly raise this difference. Later, we will show that despite this challenge in experiment design, it is possible to approximate the treatment effect from watching the video using propensity weights.

\subsection{Triggering the Experience}

The video in our example is only relevant for users that landed on the product page for X, an event known as the trigger (\cite{kohavi2009controlled}).
Users that did not land on the product page for X are irrelevant for the study, because it is not possible for the video to cause a change in sales. Including these users would attenuate the effects the business cares about, and the inability to identify these users is an imperfection of the experiment. 
Scoping the analysis to only triggered users allows experimenters to focus the impact analysis on users that would be impacted. While improving statistical power, restricting to the triggered population alters the scope of the conclusion to “Among users that viewed the product page for X, the effect of making the video available is ...”.

\subsection{Biases from Entering the Experiment}

An experiment makes inference about a broader target population, and a subset of users enter the experiment in order to gather data. We must examine how someone enters in order to tell if the sample is representative of the target. There are two important gating mechanisms: users must be on the latest version of the app, and treatment users’ apps must download a configuration file onto their device. Each has a different impact on the conclusion of the experiment. For the first, the users on the most recent version may not be representative of the target population; they may skew towards early adopters or more active users. 
%% consider removing
Therefore, results may not generalize and lack external validity. Within this skew, randomization can still partition these users equally, so scoping inference to users that entered the experiment is still internally valid.
If skew exists, then the scope of the conclusion must be qualified with “Among users that were already on the most recent version of the app...”.

The second gating mechanism induces an activation bias that is more dangerous than the first, and adds additional complexity to the intent to treat and trigger analysis.
Failing to download the configuration is a one-sided bias, because the control group does not need video configuration. This is a threat to internal validity. Additionally, the failure probability could be correlated with sales, for example both may be a function of having a strong WiFi signal. This correlation means that a comparison between the the subset of treatment users that activated, and all control users, is likely to be confounded. To avoid confounding, we must be able to identify users that entered the experiment even if they did not download the configuration; if we can only identify users after they download the configuration, then there is an activation bias. The consequence of this bias changes the scope of the question from “What is the effect of users having the ability to interact with a video on product X’s page?” to “What is the effect of enabling a user’s app to download a configuration file to enable videos on product X’s page?”.

\subsection{Scoping the Conclusion}

Finally, combining all the qualifications together, the most precise answer to the business question is either:

\begin{enumerate}
    \item “Among current users on the most recent version of the app who landed on the product page for X, the intent to treat effect of making the video available is ...”.
    \item “Among current users on the most recent version of the app, the intent to treat effect of enabling a user’s app to download video configuration is ...”.
\end{enumerate}

Designing an experiment requires several iterations. During the process, it is important to know what we can conclude, and what we cannot. The result here is internally valid for a limited target population that could have entered the experiment. Naturally, the experimenter will wonder how to generalize the result outside of the experiment. To do this, we must instrument a well structured experiment funnel.

\section{Instrumenting the Experimentation Funnel}

Keeping track of the various different scopes for the conclusion is hard. Instrumenting a funnel allows us to methodically construct what is safe to infer and what is not. The funnel contains five different stages below, and is later used to measure different effects in an unbiased way. We recommend recording all five stages. 

\includegraphics[width=\columnwidth]{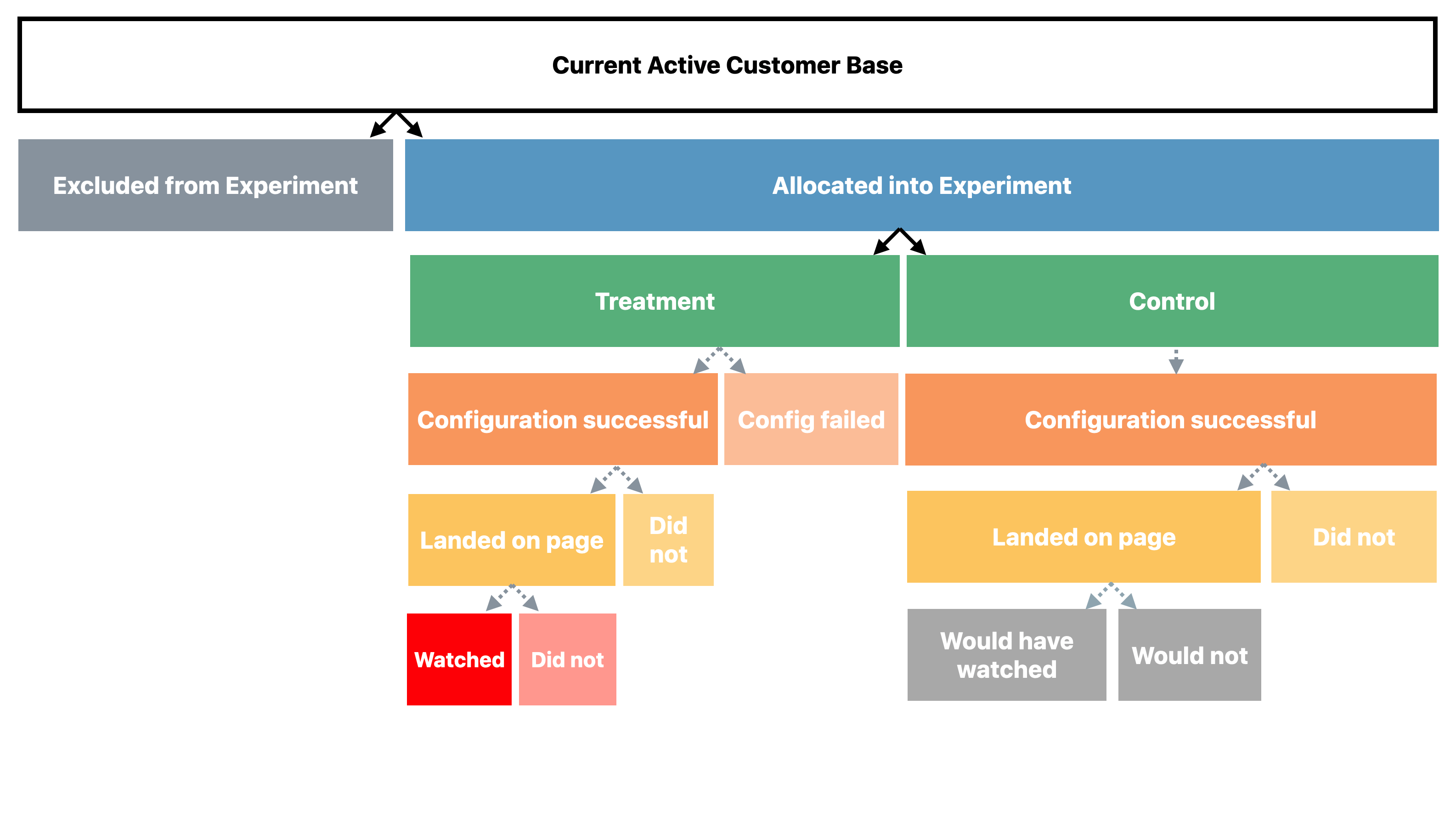}

\begin{enumerate}
    \item The \textbf{targeted} population is the pool of users the experimenters wish to study. These would be current users of the latest version of the mobile app.
    \item The \textbf{allocated} population is a subset of the targeted population, containing users who are randomly drawn to be part of an experiment. These would be current users of the latest version of the mobile app that were randomly selected to enter the experiment. It is agnostic to the treatment and control experiences. All allocated users are \textbf{assigned} into a treatment group or a control group.
    \item The \textbf{activated} population is a subset of the allocated population, containing users who have successfully received the experiment configuration appropriate for their assignment. These would be users that were able to pull the video configurations successfully.
    \item The \textbf{triggered} population is a subset of the allocated population, but does not need to be a subset of the activated population. Triggered users have have met all the criteria to be treated. These would be users that landed on the product page for X. It is agnostic to the treatment and control experiences. The app will \textbf{intend to treat} all triggered users, referencing their assignment.
    \item The \textbf{treated} population is a subset of the activated and triggered population, containing users who have experienced their assigned treatment.
\end{enumerate}

The illustration of the funnel above shows that someone who is assigned to the treatment can progress through the funnel differently than someone who is assigned to the control. Therefore, it is crucial to log all five stages in order to have a balanced comparison for the analysis of causal effects.

The analysis population shrinks going down this funnel. When planning statistical power, it should be noted that power is a function of the number of treated users, not the number of allocated users.

Using this funnel, important rates that affect the treatment effect can be calculated.

\begin{enumerate}
    \item \textbf{Allocation rate} is the percent of targeted users that allocated.
    \item \textbf{Activation rate} is the percent of allocated users that activated.
    \item \textbf{Trigger rate} is the percent of allocated users that triggered.
    \item \textbf{Compliance rate} is the percent of users who were both assigned to treatment and triggered that are treated.
\end{enumerate}

Monitoring these rates is important as they can change the result of the experiment. For example, if either the trigger rate or the compliance rate declines, then the ITT effect will be attenuated and the variance will increase, making it harder to detect an effect. If the rate fluctuates over time, it can make results that were significant become not significant. The trigger rate is also important for the practicality of the new treatment. Suppose the new video in our app generates a 10\% increase in sales among users that land on the product page for X. But only 1\% of customers land on that product page, then the total effect on sales is a more humble 0.1\%.

\section{Monitoring Imperfections}

The many hidden imperfections discussed in this paper are challenging because the experimenter may not be aware of them, and the imperfections may not be avoidable. Experimentation teams need to be prepared to balance experiment design with user experience design. One way to proactively monitor the health of the experiment design is to employ sample ratio mismatch (SRM) checks (\cite{crook2009seven}) throughout the experimentation funnel, and verify whether the funnel rates are balanced in treatment and control groups. This should be done on the allocation, activation, and trigger steps in the experimentation funnel. In particular, if SRM fails in the trigger, we should investigate whether the treatment feature has performance issues that can change user engagement. If SRM fails in the activation, we should check the size of the configuration file, which may cause failed downloads.

Within the experimentation funnel, the trigger event is the most important event because it occurs immediately before the randomized assignments go into affect. Reporting the intent to treat effect among the triggered population, including users that did not activate, is always internally valid, even when other effects need to be scrutinized. This is because the triggered population is a child of the allocated population, neither of which are unequally affected by the assignment.

\section{Adjusting Statistical Analysis}

The experiment funnel with SRM checks guarantees that the ITT effect among triggered users can be analyzed using a difference in means. While safe, inference on the triggered population is highly influenced by the trigger rates. This intent to treat effect also does not show the effect of playing the video. We can derive other types of effects using methods inspired by propensity weights (\cite{rosenbaum1983central}).

The ITT effect among users allocated into the treatment that also trigger is $\frac{1}{n_\text{T,T}} \sum_{i \in \text{T, T}} y_i - \frac{1}{n_{\text{C,T}}} \sum_{i \in \text{C, T}} y_i$ where {T,T} is the set of users assigned to the treatment that triggered, and {C,T} is the set of users assigned to the control that triggered.

We may wish to report the ITT effect for a different \textbf{reference} distribution of users, not necessarily the distribution among triggered users. Say there are two groups of users, labeled as $g_1$ and $g_2$, which are defined by a feature vector. The ratio between $g_1$ and $g_2$ users in the reference may be 1:2, but 1:10 among triggered users, caused by trigger rates of 10\% and 50\% respectively. Analysis of the triggered population does not reflect the reference population well due to the low representation of $g_1$ users. Through product changes, the business may wish to know what the ITT effect would be if the ratio of triggered $g_1$ to $g_2$ users was the same as the ratio in the allocation or target populations. Let $T(g_i)$ be the probability that an allocated user reaches the trigger, and $T'(g_i)$ be a new trigger rate. The ITT effect with a new trigger rate will be $$\frac{1}{n_\text{T, T'}} \sum_{i \in \text{T, T}} \frac{y_i T'(g_i)}{T(g_i)} - \frac{1}{n_{\text{C, T'}}}\sum_{i \in \text{C, T}} \frac{y_i T'(g_i)}{T(g_i)}.$$ In the simple case where $T'(g_i) = 1$, the ITT effect is effectively measured against the allocation distribution.

The experimenter should pay attention to the broader target population. While the funnel guides us to analyze the trigger population, the business goal is to make inference on the target population. We can repurpose the previous formula by replacing trigger rates with allocation rates.

Finally, instead of computing an ITT effect based on assignment, we may want to measure the effect of playing the video among triggered users. We compute $$\frac{1}{n_{T, T} + n_{C, T}} \sum_{i \in \text{Triggered}} \frac{w_i y_i}{p(g_i)} - \frac{(1 - w_i) y_i}{1 - p(g_i)}$$ where $w_i$ is a flag for whether a user played the video, and $p(g_i)$ is the propensity for the user to play the video.

\section{Conclusion}

We have illustrated that the ideal experiment is not always possible, even in online AB testing with its fine grained control over most aspects of the system. With our guidelines, we provide a way to communicate what is safe to conclude, and what is not. In our example, answering the desired question: “what is the impact of product videos on sales” among the target population can be restricted by selection bias and treatment compliance. When no activation bias exists, a safe and precise conclusion could be “among current users on the most recent version of the app who landed on the product page for X, the intent to treat effect of making the video available is ...”.

Awareness of imperfections and the ability to address them hinges on logging the full experiment funnel. Monitoring funnel rates with SRMs is crucial; an experiment designer cannot take it on faith that an experiment is executed perfectly. Contextualizing the broader impact of the treatment given the trigger and compliance rates also relies on the funnel. Without the funnel, activation bias can force us to draw conclusions about the effect of a user’s app being allowed to display videos rather than the effect of a user being allowed to engage with a video. By using reweighting and the funnel, we can make conclusions on the original allocation or target populations, and approximate the effect of playing the video.

We recommend experiment designers go through the exercise of discussing the target population, how users enter the experiment, how they trigger the experiment, whether the treatment is random or requires user input, and whether there are any differential biases. Afterwards, the experiment funnel and logging steps can be constructed. Having this structure will enable responsible and well scoped conclusions.

\printbibliography

\end{document}